\documentclass[
    ,final            % use final for the camera ready runs
  ]
  {aipproc}

\layoutstyle{8x11single}

\begin{document}

\title{Introducing the Dark Energy Universe Simulation Series (DEUSS)}

\classification{98.80.-k,95.36.+x,95.35.+d, 98.62.Ai,98.62.Py,98.65.Cw,98.65.Dx}
\keywords      {N-body simulations, dark energy, power spectrum, large-scale structures, quintessence, cosmology}

\author{Y. Rasera}{address={CNRS, Laboratoire Univers et Th\'eories (LUTh), UMR 8102 CNRS, Observatoire de Paris, Universit\'e Paris Diderot ; 5 Place Jules Janssen, 92190 Meudon, France}
}

\author{J-M. Alimi}{address={CNRS, Laboratoire Univers et Th\'eories (LUTh), UMR 8102 CNRS, Observatoire de Paris, Universit\'e Paris Diderot ; 5 Place Jules Janssen, 92190 Meudon, France}
}

\author{J. Courtin}{address={CNRS, Laboratoire Univers et Th\'eories (LUTh), UMR 8102 CNRS, Observatoire de Paris, Universit\'e Paris Diderot ; 5 Place Jules Janssen, 92190 Meudon, France}
}

\author{F. Roy}{address={CNRS, Laboratoire Univers et Th\'eories (LUTh), UMR 8102 CNRS, Observatoire de Paris, Universit\'e Paris Diderot ; 5 Place Jules Janssen, 92190 Meudon, France}
}

\author{P-S. Corasaniti}{address={CNRS, Laboratoire Univers et Th\'eories (LUTh), UMR 8102 CNRS, Observatoire de Paris, Universit\'e Paris Diderot ; 5 Place Jules Janssen, 92190 Meudon, France}
}

\author{A. F\"uzfa}{address={Groupe d'Application des MAth\'ematiques aux Sciences du COsmos (GAMASCO), University of Namur (FUNDP), Belgium}
}

\author{V. Boucher}{address={Center for Particle Physics and Phenomenology (CP3), Universit\'e catholique de Louvain, Chemin du Cyclotron, 2, B-1348 Louvain-la-Neuve, Belgium}
}

\begin{abstract}
In this ``Invisible Universe'' proceedings, we introduce the Dark Energy Universe Simulation Series (DEUSS) which aim at investigating the imprints of realistic dark energy models on cosmic structure formation. It represents the largest dynamical dark energy simulation suite to date in term of spatial dynamics. We first present the 3 realistic dark energy models (calibrated on latest SNIa and CMB data): $\Lambda$CDM, quintessence with Ratra-Peebles potential, and quintessence with Sugra potential.  We then isolate various contributions for non-linear matter power spectra from a series of pre-DEUSS high-resolution simulations (130 million particles). Finally, we introduce DEUSS which consist in 9 Grand Challenge runs with 1 billion particles each thus probing scales from 4 Gpc down to 3 kpc at z=0. Our goal is to make these simulations available to the community through the "Dark Energy Universe Virtual Observatory" (DEUVO), and the ``Dark Energy Universe Simulations'' (DEUS) consortium.
\end{abstract}

\maketitle

%%%%%%%%%%%%%%%%%%%%%%%%%%%%%%%%%%%%%%%%%%%%
%% MAINMATTER
%%%%%%%%%%%%%%%%%%%%%%%%%%%%%%%%%%%%%%%%%%%%
\section{Introduction}

According to current Supernova Ia Hubble diagrams (\citet{kowalski08}), and Cosmic Microwave Background anisotropies power spectrum (\citet{komatsu09}), the universe energy budget is dominated by dark energy (about 75 percents). However, the nature of dark energy is still unclear. What are the properties and origin of dark energy? How do the density and equation of state evolve across cosmic times? What is the spatial distribution?  Several physical interpretations for the observed recent acceleration of the expansion of the universe have been proposed such as cosmological constant, backreaction (\citet{buchert07}), quintessence (\citet{ratra88}) and deviations from general relativity (\citet{alimi08}) among many others. In order to discriminate between these models one has to refine current observations at the homogeneous or linear level, or to find new observables. In this article, we focus on the latter idea by investigating the imprints of dark energy on cosmic structure formation using very-high resolution N-body simulations. We note that this article is complementary to \citet{alimi09, courtin09, courtin09bis, rasera09}.

\section{Realistic quintessence models}

\begin{figure}
\begin{tabular}{rl}
 \includegraphics[width=0.6\hsize]{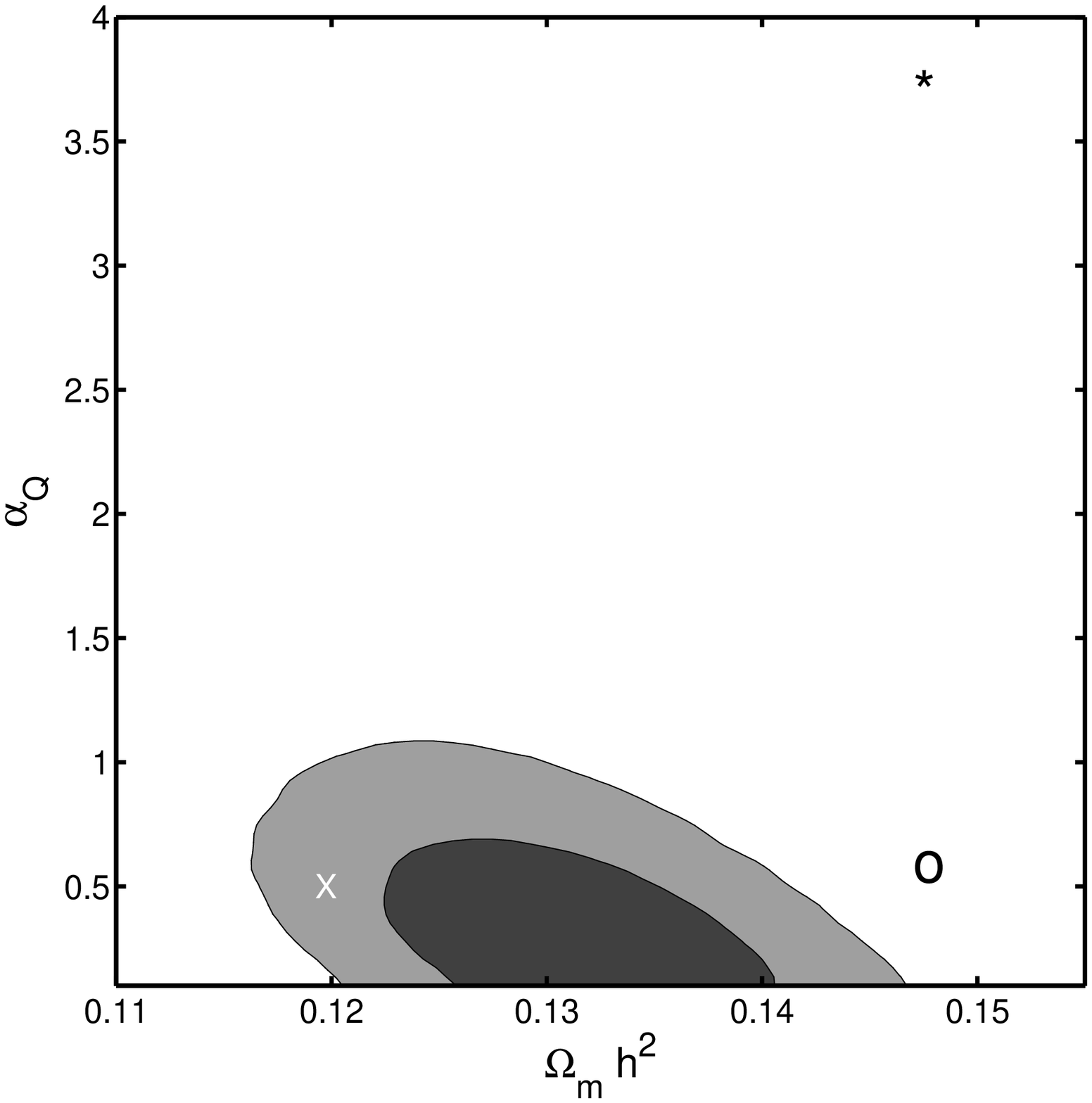} & \includegraphics[width=0.6\hsize]{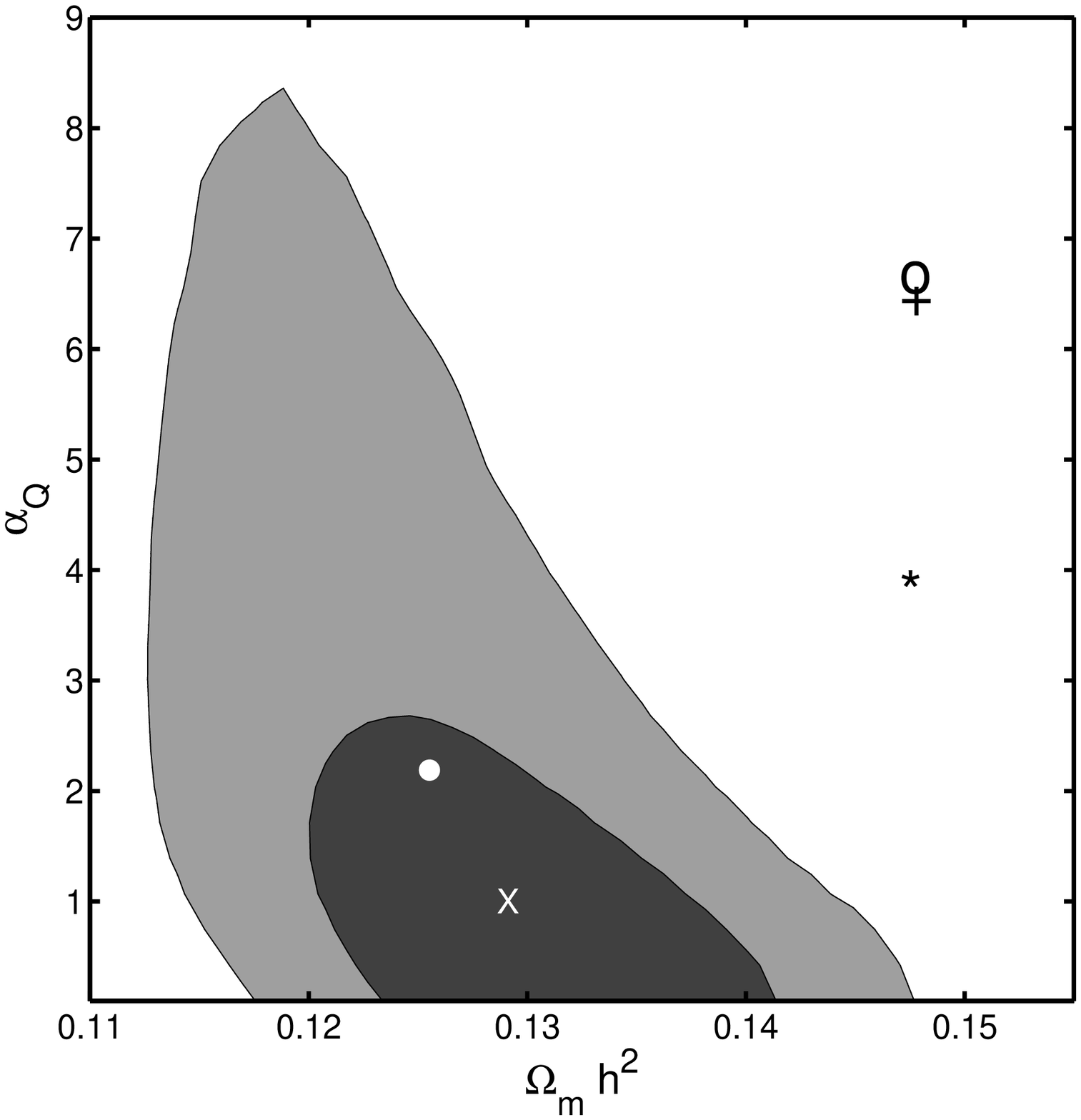}\\
\end{tabular}
\caption{$68\%$ and $95\%$ confidence regions in the $\Omega_m
  h^2-\alpha$ plane from the combined analysis
  of the UNION SN Ia Hubble diagram and WMAP-5yrs CMB data for the Ratra-Peebles quintessence
models (left) and Sugra quintessence models (right). A cross mark (X) indicates our choice for realistic RP and Sugra model
  parameters, while some other RP and Sugra model parameters assumed in the literature (for simulations) are marked
  with other symbols. These figures come from \citet{alimi09}.}
\label{conf}
\end{figure}

In this article, we consider three cosmologies with different properties: a quintessence model with a Ratra-Peebles potential (\citet{ratra88}), a quintessence model with supergravity corrections (\citet{brax00}) and the standard $\Lambda$CDM model. We however expect our physical conclusions to apply on a much wider class of models. Most of the previous N-body simulations with dark energy assume $\Lambda$CDM cosmological parameters and even $\Lambda$CDM linear power spectrum for setting initial conditions (see however recent papers from \citet{casarini09,jennings09}). To overcome these issues, we create a set of two ``realistic quintessence models'' plus a reference $\Lambda$CDM WMAP-5yrs cosmology. We modify CAMB to take into account the fluctuations of quintessence in order to compute the linear matter  power spectrum. We then perform a likelihood analysis of the UNION SN Ia Hubble diagram and WMAP-5yrs CMB data to select viable quintessence models as illustrated Fig.~\ref{conf}. We choose $\Omega_m=0.23$, $\alpha=0.5$, $\sigma_8=0.66$ for Ratra-Peebles quintessence models, $\Omega_m=0.25$, $\alpha=1$, $\sigma_8=0.73$ for Sugra, $\Omega_m=0.26$, $\sigma_8=0.79$ for $\Lambda$CDM. It is worth noting that quintessence models favor lower $\Omega_m$ and $\sigma_8$. Our three models are degenerated regarding to CMB and SNIa constraints and one needs additionnal constraints (such as from the non-linear regime) in order to discriminate them.

\section{High-resolution N-body simulations}

\begin{figure}

\begin{tabular}{cc}
\includegraphics[width=0.5\hsize]{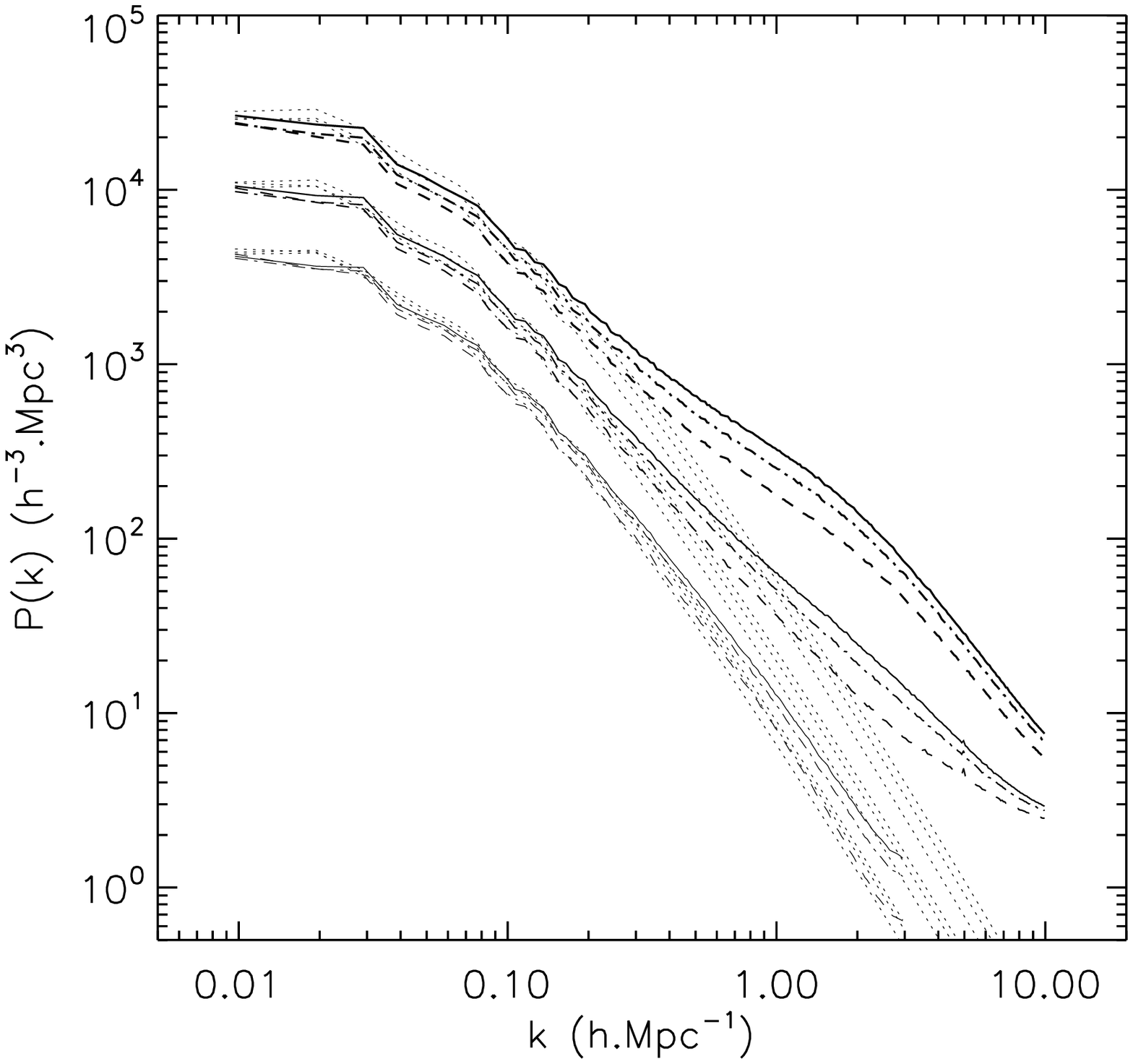}&\includegraphics[width=0.5\hsize]{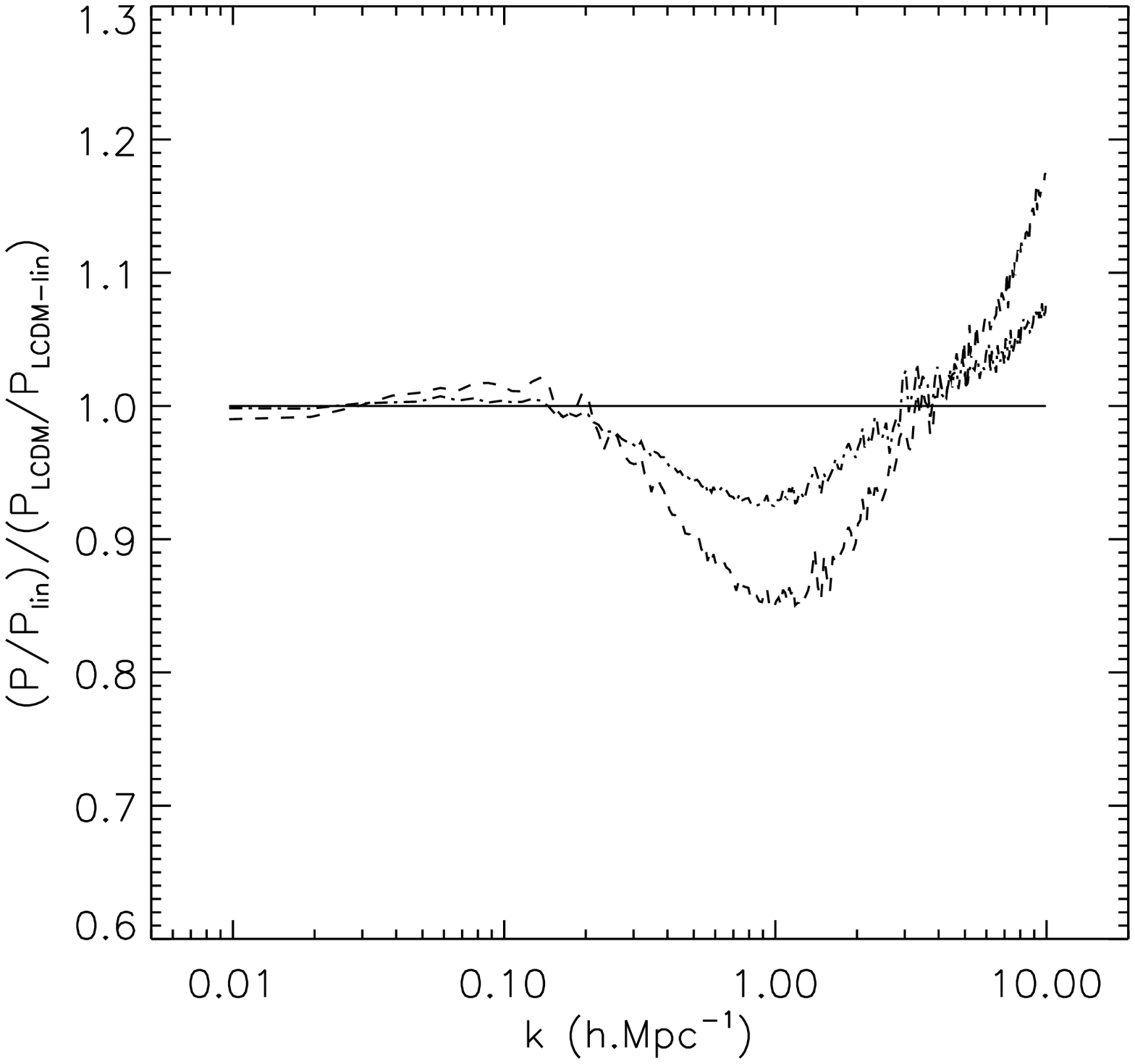}\\
\includegraphics[width=0.5\hsize]{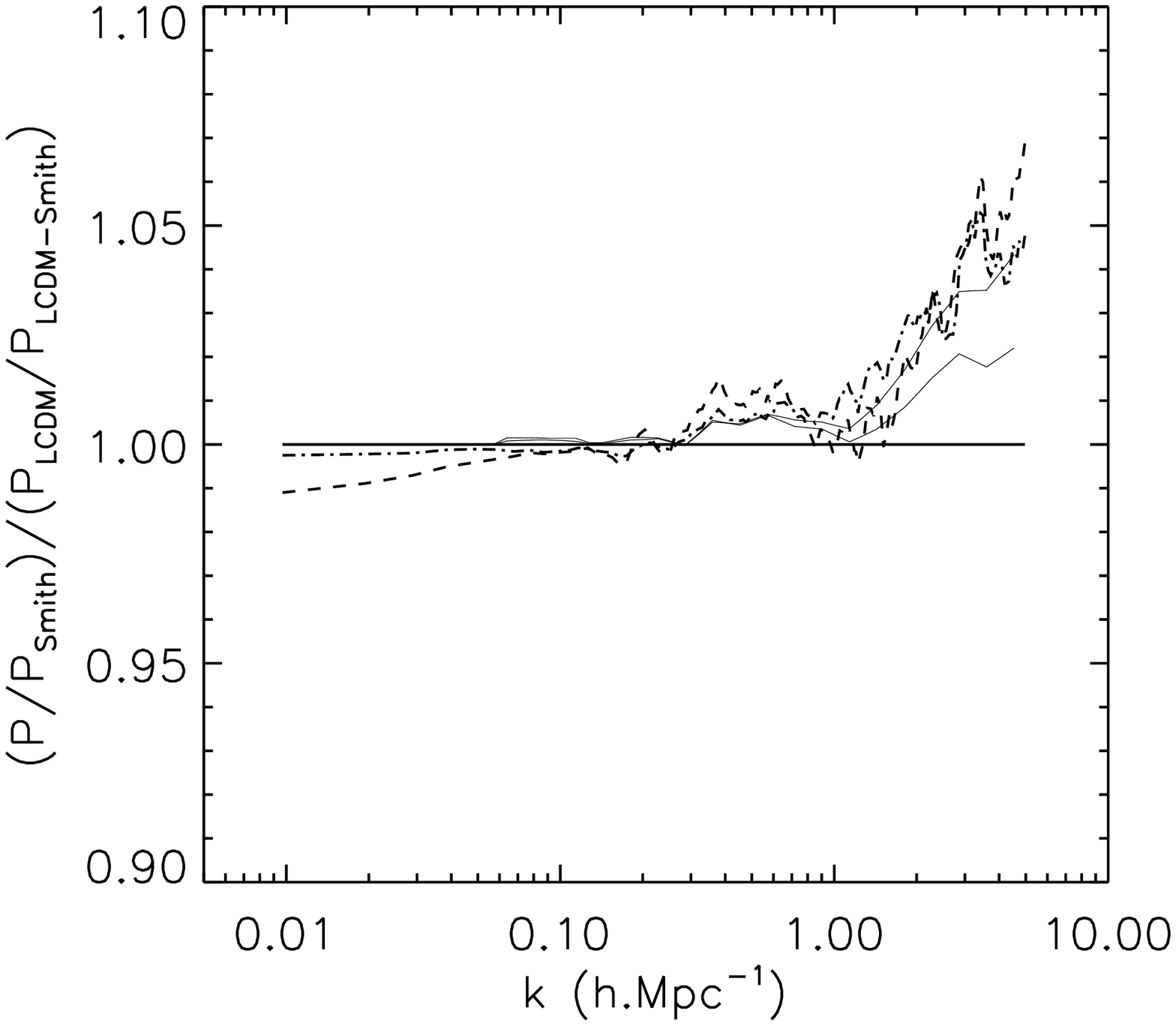}&\includegraphics[width=0.5\hsize]{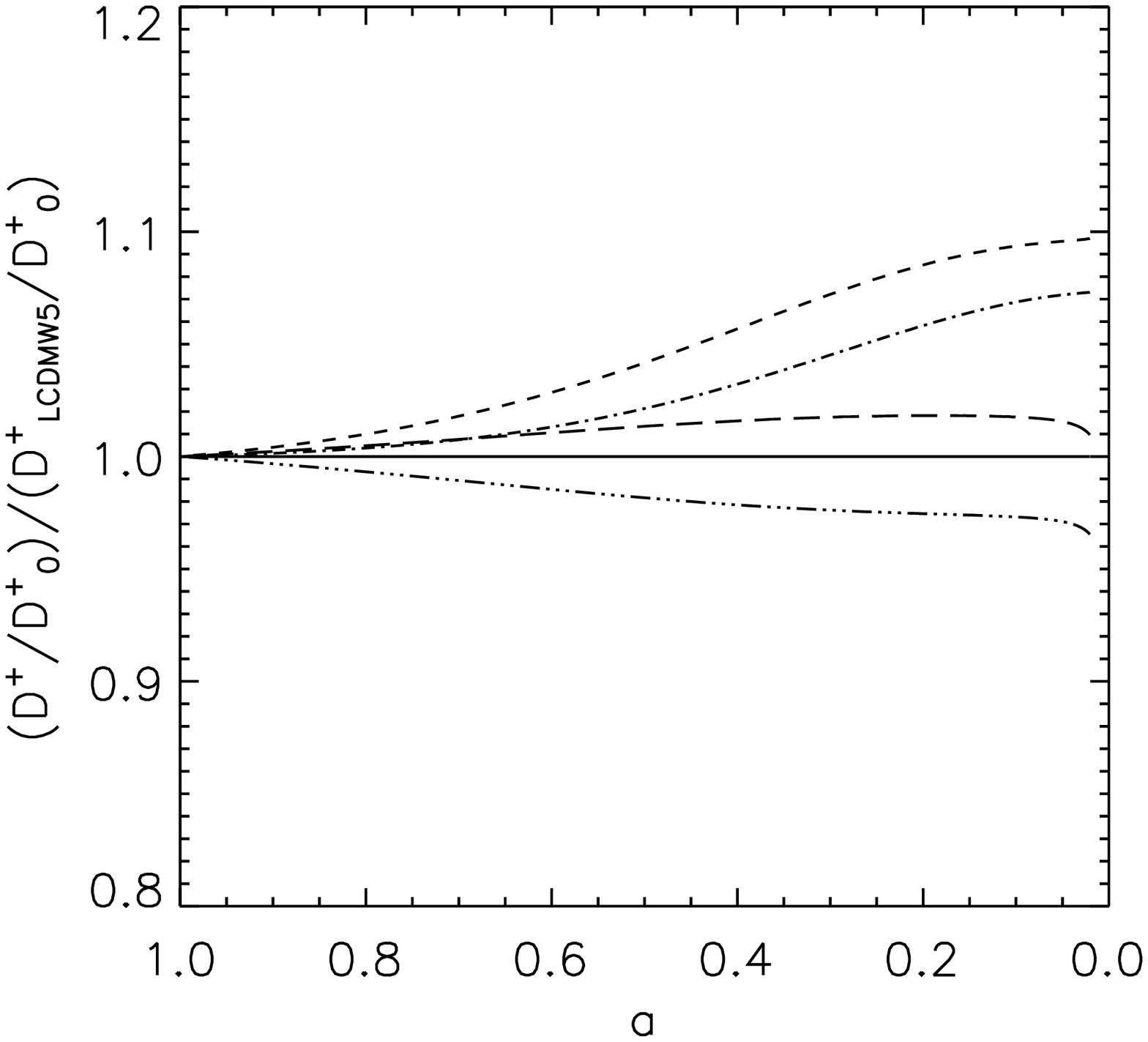}\\
\end{tabular}
\caption{In all the plots $\Lambda$CDM is the continuous line, Ratra-Peebles the dashed line, and Sugra the dot-dashed line. Top left panel: power spectrum at z=0 (upper triplet), 1 and 2.33 (lower tripplet) and associated linear power spectra (dotted lines). Top right panel: Ratio of power spectrum (normalized by the linear power spectrum) to the $\Lambda$CDM power spectrum. Bottom left panel: Ratio of the non-linear power spectrum (normalized by the \citet{smith03} fit) relative to the $\Lambda$CDM one. The thin continuous line are the constant eos expectation for RPCDM (top) and SUCDM (bottom).  Bottom right panel: Linear growth factor histories relative to the $\Lambda$CDM WMAP5 one (for comparison $\Lambda$CDM WMAPI and $\Lambda$CDM WMAPIII are shown as triple dot dashed line and long dashed line respectively). The deviations at high-k of the power spectra are correlated with the linear growth histories.These figures come from \citet{alimi09}}
 \label{history}

\end{figure}

We run a series of 9 high resolutions simulations with $512^3$ particles each, for the three ``realistic'' cosmologies described above and for three box lengthes ($162$, $648$ and $1296$~h$^{-1}$Mpc). Initial conditions are set using the MPGRAFIC software (\citet{prunet08}). Particles evolution is computed with the Adptative Mesh Refinement (AMR) code RAMSES (\citet{teyssier02,rasera06}). Finally, power spectra are computed using POWMES (\citet{colombi09}). Resulting matter power spectra are shown in Fig.~\ref{history}. From the power spectra plot (top-left), it can be seen that quintessence universes are less structured than $\Lambda$CDM ones at all redshifts. A large part of the discrepancies is imputed to the different linear power spectra (lower $\sigma_8$ but also quintessence-specific shape). In order to remove this linear contribution, we show (on the top-right plot) ratios of quintessence models power spectra (normalized by the linear one) to $\Lambda$CDM power spectrum (also normalized by the linear one) at z=0. As expected for small wavenumbers (below 0.1~h/Mpc), the three power spectra superimpose. However, differences arise above 0.1~h/Mpc, which are mostly caused by non-linear amplification of the linear growth rate. This effect is well accounted for by the \citet{smith03} fitting function. The bottom-left plot confirms this point for wavenumbers below 1~h/Mpc. However, some important discrepancies remain for k larger than 1~h/Mpc. Using a constant equation of state $\omega=\omega_0$ and the fit of \citet{mcdonald06} doesn't solve the issue. It indicates that each dark energy model leaves a specific imprints on the high-k tail of the power spectrum, and that using constant equation of state approximation (\citet{ma99,mcdonald06,lawrence09}) results in several percents errors on the power spectrum. The correlation with the linear growth history (bottom-right plot, see also \citet{ma07}) indicates that the non-linear power spectrum records the history of structure formation (\citet{alimi09}). This has yet to be properly accounted for in analytical predictions since most of the current fits (\citet{peacock96,smith03}) only depend on the instantaneous value of the linear growth rate.

\section{Grand Challenge runs: DEUSS}

\begin{table}
\begin{tabular}{cccc}\hline
Parameters  & $162\rm\; h^{-1}Mpc$ & $648\rm\; h^{-1}Mpc$ & $2592\rm\; h^{-1}Mpc$\\
\hline\hline
\textbf{$\Lambda$CDM W5}&&&\\
$z_{\rm i }$ & $137$ & $93$ & $56$ \\
$m_p \rm (h^{-1}\;M_\odot)$ & $2.86\times 10^8$ & $1.83\times 10^{10}$ & $1.17\times 10^{12}$ \\
$\Delta_x \rm (h^{-1} kpc)$ & $2.47$ & $9.89$ & $39.6$ \\
\hline
\textbf{Ratra-Peebles}&&&\\
$z_{\rm i }$ & $118$ & $81$ & $50$ \\
$m_p \rm (h^{-1}\;M_\odot)$ & $2.53\times 10^8$ & $1.62\times 10^{10}$ & $1.04\times 10^{12}$ \\
$\Delta_x \rm (h^{-1} kpc)$ & $2.47$ & $9.89$ & $39.6$ \\
\hline
\textbf{Sugra}&&&\\
$z_{\rm i }$ & $134$ & $92$ & $56$ \\
$m_p \rm (h^{-1}\;M_\odot)$ & $2.75\times 10^8$ & $1.76\times 10^{10}$ & $1.13\times 10^{12}$ \\
$\Delta_x \rm (h^{-1} kpc)$ & $2.47$ & $9.89$ & $39.6$ \\
\hline
\end{tabular}
\caption{Simulation parameters for the DEUSS Grand Challenge runs at the IDRIS supercomputing center. Each simulation contains $1024^3$ particles
with $1024^3$ coarse-grid cells, 6 refinement levels and has been evolved down to $z=0$. The various entries report the values of the simulation initial redshift
  $z_{\rm i}$, particle mass $m_p$ and spatial resolution $\Delta_x$.}
\label{tab3}
\end{table}

\begin{figure*}
\includegraphics[width=\hsize]{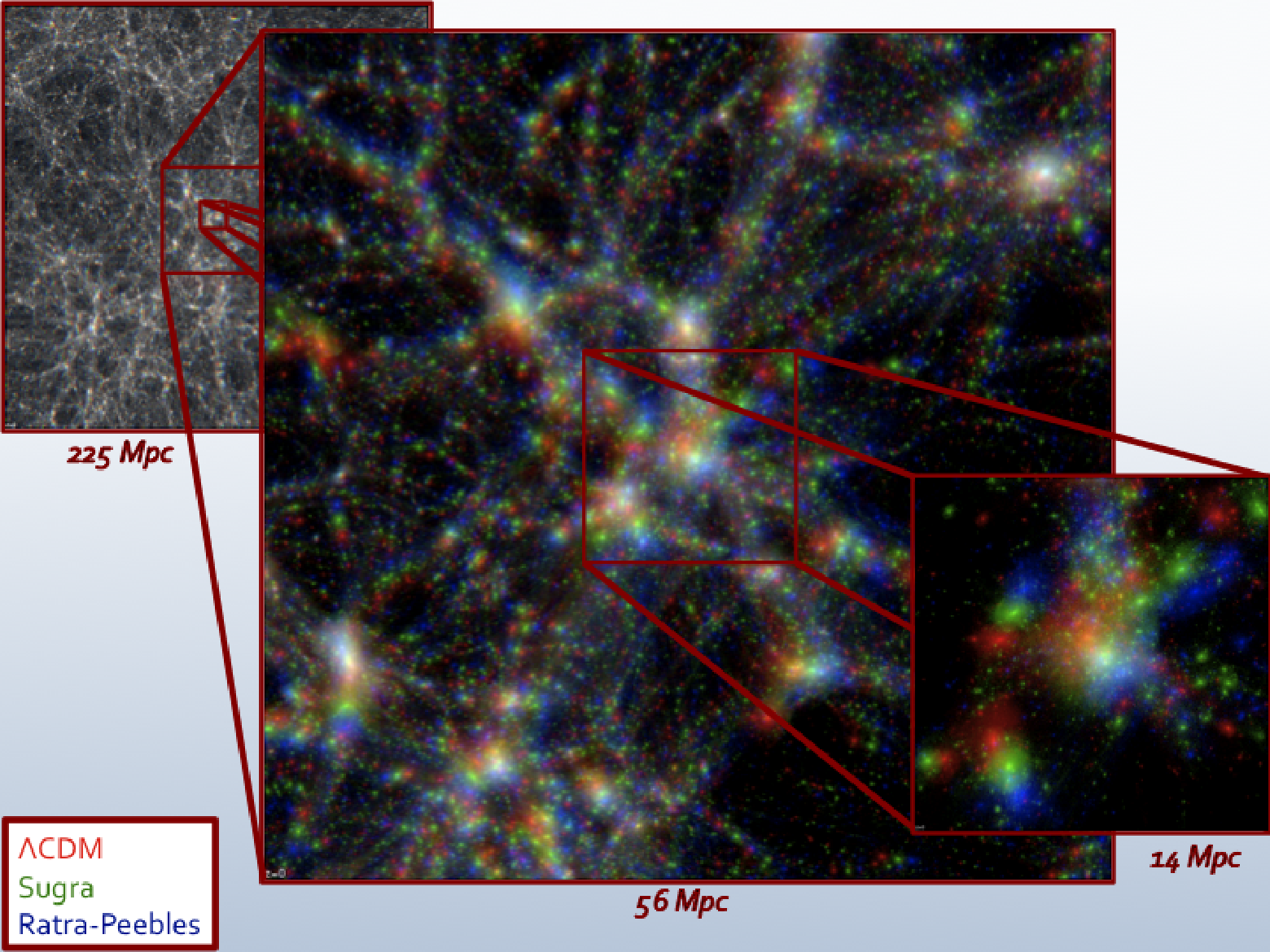}
\caption{Projected density maps from the Dark Energy Universe Simulation Series (DEUSS) at z=0 for three cosmologies: $\Lambda$CDM (red), Sugra (green) and Ratra-Peebles (blue). The spatial dynamics is illustrated by this zooming sequence from the full box scale (162~Mpc/h) onto one halo (at 10~Mpc/h scale) among the 300000 well-resolved halos in the simulation volume. At large scale, the picture is essentially black and white because differences between cosmologies are small. However, the imprints of dark energy (multicolor pictures) are clearly visible on non-linear scales. Overall, DEUSS consist in 9 Grand Challenge simulations with 1 billion particles resolving scale from 4~Gpc to 3~kpc. It has been run on 4096 Blue Gene/P cores at the IDRIS supercomputing center, using 5000000 mono-cpu hours thanks to Horizon Project at LUTH. This figure comes from http://www.idris.fr/docs/docu/projets-Babel/DEUSS/CR-projet-DEUSS.html.}
\end{figure*}

In order to gain further physical insight into dark energy's imprints on cosmic structure formation, one needs to investigate its signature at all scales: from galaxies inner profiles to Hubble volume cluster counts. To this aim, we perform the unprecedented Dark Energy Universe Simulation Series (DEUSS). It consists in 9 Grand Challenge simulations with 1 billion particles each (and up to 7 billions cells), resolving scales from 4~Gpc down to 3~kpc for the three considered cosmologies: $\Lambda$CDM, Sugra and Ratra-Peebles. Taking advantage of its Peano-Hilbert domain decomposition, the MPI code RAMSES has been run on 4096 Blue Gene/P cores at the IDRIS supercomputing center (last trimester 2008). It represents 5 millions mono-cpu hours (that is to say about 600 years) of computationnal time that have been obtained by the Horizon Project at LUTH. The run parameters are described in Table~\ref{tab3}: simulations box lengthes are 162~Mpc/h, 648~Mpc/h and 2592~Mpc/h.

We backup three kinds of data during the runs. First, we save the masses, positions, velocities and identities of all particles for 24 snapshots for each simulation (expansion factor $a=0.025$, $0.05$, $0.1$, $0.15$, $0.2$, $0.25$, $0.3$, $0.35$, $0.4$, $0.45$, $0.5$, $0.55$, $0.6$, $0.65$, $0.7$, $0.729$, $0.75$, $0.8$, $0.823$, $0.85$, $0.9$, $0.95$ and $1$), that is to say a total of 216 snapshots. Second, we construct 9 lightcones around an observer at the center of the simulation by dumping particles from the light-travel distant slice at each coarse time step. For the largest box, the lightcones cover $4\pi$ steradian up to z=1 whereas for the smallest volume, the opening angle is 10 degrees by 10 degrees. Third, we save 6 samples of particles at every coarse time step in a parallelepiped of size $1/16 \times 1/16 \times 1$ times the full volume of the simulations. The largest number of coarse time steps is 1419. The total amount of data is about 40~To.

Unfortunately, the output files are organized along the Peano-Hilbert curve which could be non-trivial to manipulate. They also contain a lot of double-precision and not-useful information (except for restarting the run). We therefore run a parallel Slicer in order to split the data in $8^3$ cubic regions or ``fields'' and to keep only useful simple precision data. We also develop and run a parallel Friend-of-Friend halo finder (linking length b=0.2) on 512 processes. We detect all halos above 100 particles in half of the outputs (the analysis of the others is on-going). We then sort the particles halo per halo and backup these 512 ``halo'' files. Overall halo masses range from $3\times 10^{10}$~Mpc/h to $8\times 10^{15}$~Mpc/h, there are up to 500 000 halos per snapshot and up to 3 millions particles per halos. The total number of analyzed halos is about 20 millions (20 other millions will be detected soon). With this two kinds of files (halos and fields), most of subsequent post-processing can be serial. Moreover, data are user-friendly to manipulate.

As an illustration of the results we show a composite view of the density field for the 3 simulations of box length 162 Mpc/h at z=0 with color coding for each cosmology: $\Lambda$CDM (red), Sugra (green) and Ratra-Peebles (blue). On large scale, the map is essentially black and white indicating little difference between cosmologies whereas on smaller scales (40 and 10~Mpc/h) the differences between the models show up (multi-color filaments and halos). Discrepancies are therefore amplified by the non-linear regime of structure formation and can help to break degeneracies between cosmologies.

Our plan is to share these data with the ``dark energy'' community. To this aim we are starting the DEUS (Dark Energy Universe Simulations) consortium. The goal is to gather observers and theoreticians in order to analyse the data, prepare future dark energy observational missions (weak lensing, strong lensing, BAO, cluster counts, etc.), improve analytical predictions (power spectrum, mass functions, PDF, density profiles, etc.), and better understand the imprints of dark energy on structure formation. We are also building a database to diffuse the results. The database should contain all the properties of all halos and fields from the different outputs. The strength of this Dark Energy Universe Virtual Observatory (DEUVO) is that it follows the IVOA data model to allow future interoperability. A website is also currently under development (http://www.deus-consortium.org).

\section{Conclusion}

This article can be summarized as follow.
\begin{enumerate}
\item Unlike most of the previous numerical simulations, we consider three realistic dark energy models. These models are calibrated on latest SNIa and CMB data and take into account quintessence clustering in the linear regime. As a result, $\Omega_m$ and $\sigma_8$ are lower for quintessence cosmologies.
\item We run a series of pre-DEUSS high-resolution runs (130 millions particles). We distinguish three dark energy signatures on the non-linear power-spectra: linear normalisation and shape, non-linear amplification, and a record of the past history of cosmic structure formation (which is not accounted for in most of current fits). 
\item We run the unprecedented DEUSS Grand Challenge runs in order to probe structure formation at higher resolution and on larger scale than before (ie. from 3~kpc to 4~Gpc) for the three realistic dark energy models. The simulations evolve 1 billion particles and up to 7 billions AMR cells. Our data consist in 216 snapshots, 9 lightcones and 6 samples. We have already analyzed half of the snapshots and produced a catalog of 20 millions halos and 60 thousands cubic fields. These user-friendly data represents a unique way for observers and theoreticians from the (newly-created) DEUS consortium to seek for new observables using the DEUVO database (in construction). It will help us to break degeneracies between dark energy models and to improve our understanding of the role of dark energy on cosmic structure formation.
\end{enumerate}

%%%%%%%%%%%%%%%%%%%%%%%%%%%%%%%%%%%%%%%%%%%%
%% Sample figure:
%%
%% The option [height=...] scales the picture to the given height,
%% without it it would be printed at its nominal size
%%%%%%%%%%%%%%%%%%%%%%%%%%%%%%%%%%%%%%%%%%%%

%\begin{figure}
  %\includegraphics[height=.3\textheight]{golfer}
%  \caption{Picture to fixed height}
%\end{figure}

%%%%%%%%%%%%%%%%%%%%%%%%%%%%%%%%%%%%%%%%%%%%%%%%
%% BACKMATTER
%%%%%%%%%%%%%%%%%%%%%%%%%%%%%%%%%%%%%%%%%%%%%%%%

\begin{theacknowledgments}
 This work was granted access to the HPC resources of CCRT and IDRIS under the
allocations 2008-i20080412287/i2009042287 and 2009-t20080412191/x2009042191 made by GENCI (Grand Equipement National de Calcul Intensif). We would like to thank Romain Teyssier, Simon Prunet, Stephane Colombi, Philippe Wautelet and all the IDRIS engineers for their great help in running DEUSS.
\end{theacknowledgments}

%%%%%%%%%%%%%%%%%%%%%%%%%%%%%%%%%%%%%%%%%%%%%%%%
%% The bibliography can be prepared using the BibTeX program or
%% manually.
%%
%% The code below assumes that BibTeX is used.  If the bibliography is
%% produced without BibTeX comment out the following lines and see the
%% aipguide.pdf for further information.
%%
%% For your convenience a manually coded example is appended
%% after the \end{document}
%%%%%%%%%%%%%%%%%%%%%%%%%%%%%%%%%%%%%%%%%%%%%%%%


\begin{thebibliography}{9}
\bibitem[Alimi et al.(2010)]{alimi09} Alimi, J.-M., F{\"u}zfa, A., Boucher, V., Rasera, Y., Courtin, J., \& Corasaniti, P.-S.\ 2010, MNRAS, 401, 775 
\bibitem[Alimi \& Fuzfa(2008)]{alimi08} Alimi, J.-M., Fuzfa, A.\ 2008, Journal of Cosmology and Astro-Particle Physics, 9, 14
\bibitem[Brax \& Martin(2000)]{brax00} Brax, P., \& Martin, J.\ 2000, Phys. Rev. D, 61, 103502 
\bibitem[Buchert(2007)]{buchert07} Buchert, T.\ 2007, XIIth Brazilian School of Cosmology and Gravitation, 910, 361 
\bibitem[Casarini et al.(2009)]{casarini09} Casarini, L., Macci{\`o}, A.~V., \& Bonometto, S.~A.\ 2009, Journal of Cosmology and Astro-Particle Physics, 3, 14 
\bibitem[Colombi et al.(2009)]{colombi09} Colombi, S., Jaffe, A., Novikov, D., \& Pichon, C.\ 2009, MNRAS, 393, 511 
\bibitem[Courtin et al.(2010)]{courtin09} Courtin, J., Rasera, Y., Alimi, J.~-., Corasaniti, P.~-., Boucher, V., \& Fuzfa, A.\ 2010, arXiv:1001.3425 
\bibitem[Courtin et al.(2010b)]{courtin09bis} Courtin, J., Alimi, J-M., Rasera, Y., Corasaniti, P-S., F\"uzfa, A., Boucher, V., 2010, ``Invisible Universe conference'',in press.
\bibitem[Jennings et al.(2009)]{jennings09} Jennings, E., Baugh, C.~M., Angulo, R.~E., \& Pascoli, S.\ 2009, arXiv:0908.1394 
\bibitem[Komatsu et al.(2009)]{komatsu09} Komatsu, E., et al.\ 2009, ApJs, 180, 330 
\bibitem[Kowalski et al.(2008)]{kowalski08} Kowalski, M., et al.\ 2008, ApJ, 686, 749 
\bibitem[Lawrence et al.(2009)]{lawrence09} Lawrence, E., Heitmann, K., White, M., Higdon, D., Wagner, C., Habib, S., \& Williams, B.\ 2009, arXiv:0912.4490 

\bibitem[Ma et al.(1999)]{ma99} Ma, C.-P., Caldwell, R.~R., Bode, P., \& Wang, L.\ 1999, ApJL, 521, L1 
\bibitem[Ma(2007)]{ma07} Ma, Z.\ 2007, ApJ, 665, 887 
\bibitem[McDonald et al.(2006)]{mcdonald06} McDonald, P., Trac, H., \& Contaldi, C.\ 2006, MNRAS, 366, 547 
\bibitem[Peacock \& Dodds(1996)]{peacock96} Peacock, J.~A., \& Dodds, S.~J.\ 1996, MNRAS, 280, L19 
\bibitem[Prunet et al.(2008)]{prunet08} Prunet, S., Pichon, C., Aubert, D., Pogosyan, D., Teyssier, R., \& Gottloeber, S.\ 2008, ApJs, 178, 179 
\bibitem[Ratra \& Peebles(1988)]{ratra88} Ratra, B., \& Peebles, P.~J.~E.\ 1988, Phys. Rev. D, 37, 3406 
\bibitem[Rasera \& Teyssier(2006)]{rasera06} Rasera, Y., \& Teyssier, R.\ 2006, A\&A, 445, 1 
\bibitem[Rasera et al.(2010)]{rasera09} Rasera, Y., Courtin, J., Alimi, J-M., Corasaniti, P-S., F\"uzfa, A., Boucher, V., 2010, MNRAS, in prep
\bibitem[Smith et al.(2003)]{smith03} Smith, R.~E., et al.\ 2003, MNRAS, 341, 1311 
\bibitem[Teyssier(2002)]{teyssier02} Teyssier, R.\ 2002, A\&A, 385, 337 
\end{thebibliography}
\end{document}